\documentclass[twocolumn,showpacs,preprintnumbers,amsmath,amssymb]{revtex4}
\pdfoutput=1 

\usepackage{amsmath,amssymb,bbm,graphicx,color}

\newcommand{\be}{\begin{equation}}
\newcommand{\ee}{\end{equation}}
\newcommand{\ba}{\begin{array}}
\newcommand{\ea}{\end{array}}
\newcommand{\bqa}{\begin{eqnarray}}
\newcommand{\eqa}{\end{eqnarray}}

\newcommand{\tr}{\mbox{Tr}}

\newcommand{\ket}[1]{\ensuremath{| #1 \rangle}}

\newcommand{\ie}{{\it i.e.}}

\begin{document}

\title[Optimal dynamical control of many-body entanglement]{Optimal dynamical control of many-body entanglement}

\author{Felix Platzer, Florian Mintert, Andreas Buchleitner}
\affiliation{
Institute of Physics,
Albert-Ludwigs University of Freiburg,\nolinebreak[4]
Hermann-Herder-Str. 3,
79104 Freiburg
Germany}

\date{\today}

\begin{abstract}
The preparation of highly entangled many-body systems is one of the central challenges of both basic and applied science. The complexity of interparticle interaction and environment coupling increases rapidly with the number of to-be-entangled subsystems, rendering the requirements on the control of many-body quantum systems ever more restrictive. We propose an approach that allows to derive optimal control Hamiltonians in a purely algebraic fashion. These drive a composite quantum system rapidly into that highly entangled state which can be created most efficiently for a given interaction mechanism, and which bears entanglement that is robust against decoherence.
 \end{abstract}
 
\maketitle

Quantum entanglement is a genuine trait of quantum mechanics without classical analogue. Whereas its central role as a resource for any task of quantum computation and quantum communication is undoubted, its relevance even
in seemingly classical bio-molecules is under active, controversial debate \cite{briegel_bio,arxiv09053787}.
Quantum correlations, {\it i.e.} entanglement naturally arises between interacting particles.
But, at the same time, the coupling of the particles to a large number of environmental degrees of freedom degrades quantum coherences that give rise to non-classical correlations. Thus, the interplay of inter-particle coupling and decoherence determines the time evolution of entanglement, and, generally, limitations to the life-time of entanglement due to environment coupling are ever more restrictive for systems of growing size. Therefore, we need tools to actively steer many particle quantum systems into highly entangled states, which ideally should be robust against environmental influences.
Only with such tools at hand will we be able to process quantum information beyond proof of principle experiments.
Furthermore, this endeavor will improve our understanding of entanglement evolution in large, open quantum systems, and of its possible role for biological functionality \cite{HohjaiLee06082007}.

In the present paper, we derive coherent control techniques that allow to exploit inter-particle interactions in an optimal fashion and, simultaneously, reduce the influence of decoherence, in order to create entanglement rapidly, and to slow down its decay.
Coherent control, {\it i.e.} the guidance of a system by an externally applied, time dependent control Hamiltonian $H_\text c (t)$, has proven very successful for optimal gate-implementation \cite{glaserGRAPE} and for the suppression of decoherence in noisy environments \cite{controlcohrev, lukcluster}, for example by restricting the time evolution to decoherence-free subspaces of the Hilbert space \cite{lidardecohfree}, or by dynamically decoupling the system from the environment~\cite{dyndecoup_nature}.
Typically optimal control is based on a target functional that is to be maximized by a suitably constructed control Hamiltonian.
The generic choice for this functional is fidelity, {\it i.e.} the success probability of reaching a predefined target state \cite{GlaserEnt}.
However, since entanglement is a non-local property, it is invariant under local unitary transformations, so that there is a multitude of states with equivalent entanglement properties,
that generally require different resources for their production and have different entanglement decay times \cite{borplasrob}.
Therefore the restriction to specific target states is disadvantageous when entanglement is the figure of merit.
In this context, an entanglement measure as target functional is more favorable,
and its use very naturally results in the creation of the state that
is best adapted to the system properties --
{\it i.e.} that is most easily accessible for a given interaction,
and that has the most robust entanglement properties under the specific environment coupling.

Entanglement measures have already been employed as target functionals for two-qubit systems, in the context of gate optimization \cite{entgates},
or for the optimal choice of an initial state to evolve into a strongly entangled state \cite{loccon}.
But an extension to higher dimensional, multipartite systems is a highly nontrivial step, since the quantification of mixed state entanglement is excessively elaborate.
Owing to the fact that dynamical optimization of entanglement implies its repeated evaluation, an absolutely necessary prerequisite for the usage of an entanglement measure as target functional is the possibility to evaluate it efficiently.
This requirement is met by an algebraic lower bound of the multipartite concurrence $c(\varrho)$ \cite{woocon,aolita:022308}, which reads
\bqa
\label{eq:targetfunc}
&\tau(\varrho) &=\tr\, \varrho\otimes\varrho \, {\bf A} \leq c(\varrho)^2 \ ,\mbox{ with}\\
&&{\bf A}=4\Big({\bf P}_{+}-P^{1}_{+}\otimes\hdots\otimes P^{N}_{+}-(1-2^{1-N}){\bf P}_{-}\Big)\nonumber\ ,
\eqa
where ${\bf P}_{\pm}$ are projectors onto the symmetric and antisymmetric subspace of the duplicate Hilbert space ${\cal H}\otimes{\cal H}$, and $P_{+}^{i}$ ($i=1,\hdots,N$) is the projector onto the symmetric subspace of the duplicate Hilbert space ${\cal H}_i\otimes{\cal H}_i$ of subsystem $i$.
Like any entanglement measure, $\tau$ is invariant under local unitaries \cite{PhysRevA.58.1833}, and it is a valid lower bound for the convex roof construction \cite{PhysRevA.62.032307} for mixed states---\ie\ the minimal average entanglement over all pure state decompositions of $\varrho$.
This bound is particularly tight for weakly mixed states \cite{flomix},
and exact for pure states, what is precisely the class of states that we are aiming at when trying to reach high entanglement.

The feasibility of any control scheme depends on the range of control Hamiltonians that can be experimentally engineered.
Since, in general, one cannot assume that interactions can be engineered, we presume that control induces only single particle dynamics---\textit{i.e.} we consider local control.
Thus, we can phrase our general control strategy as follows: which local rotations do we need to apply, such that the effect of environment is least detrimental---or even beneficial---and such that interactions are optimal for the creation entanglement?

A very popular control strategy that underlies algorithms such as Krotov's algorithm \cite{tannorkrotov} and GRAPE \cite{glaserGRAPE} is based on variational methods, in which optimal control pulses are reached through many repeated forward and backward propagations of the target functional.
In contrast, our strategy is based on the optimization of temporal derivatives of $\tau$, and requires only one single propagation. The underlying idea is that a maximization of $\dot \tau$ will result in a temporal increase of $\tau$. However, $\dot\tau$ is independent of {\em local} control fields because of the invariance of $\tau$ under the induced local unitary dynamics. Therefore, we will resort to the maximization of the curvature $\ddot \tau$, which in the course of time will result in an increase of $\dot\tau$, which, in turn, yields the desired increase of $\tau$.

Both $\dot\tau$ and $\ddot\tau$ depend not only on $\varrho$,
but also on its time-derivatives
\be
\dot \tau = 2\, \tr\, \dot \varrho \otimes \varrho \, {\bf A}\ ,\hspace{.5cm}
 \ddot \tau = 2\, \tr\, (\ddot \varrho \otimes \varrho + \dot \varrho \otimes \dot \varrho) \, {\bf A}\ ,
 \ee
the first of which is governed by the master equation \cite{Lindblad}
\begin{equation}
\dot \varrho = \frac i \hbar [ H_{\text{sys}} + H_{\text{c}}, \varrho ] + \underbrace{\sum_i \frac {\Gamma_i} 4 (2\mu_i \varrho \mu_i^\dagger - \varrho \mu_i^\dagger \mu_i -  \mu_i^\dagger \mu_i \varrho)}_{{\cal L}(\varrho)}\ .
\label{eq:drho}
\end{equation}
The second derivative $\ddot\varrho$ is obtained through iteration of Eq.~\eqref{eq:drho}.
The Hamiltonian in the coherent part of this master equation is the sum of the local control Hamiltonian $H_\text{c}$, and of the system Hamiltonian $H_{\text{sys}}$, involving interactions between subsystems, which are assumed to be inaccessible to control.
The incoherent part ${\cal L}(\varrho)$ in Eq.~\eqref{eq:drho} arises from the system's interaction with its environment, and typically leads to  entanglement decay \cite{Briegel, flodeco}.

Although the control Hamiltonian $H_\text{c}$ enters at second order
via $\ddot\varrho$ and via $\dot\varrho \otimes \dot\varrho$, the curvature $\ddot\tau$ is actually a linear function of $H_\text{c}$,
since the quadratic terms in $H_\text{c}$ correspond to local unitary dynamics which leave $\tau$ unchanged.
However, linear terms describe the interplay $H_{\text{c}}$ with the inter-particle interactions $H_{\text{sys}}$, and with the incoherent dynamics ${\cal L}(\varrho)$.
Since this results in global or non-unitary dynamics, respectively, $\ddot\tau$ actually {\em does} depend on $H_\text{c}$.
Physically, this reflects that local control induces an evolution towards a state with different dynamical properties.
And, even though entanglement can {\em not} be changed through this local driving alone, the induced changes in the state's dynamical properties can influence the impact of interaction and decoherence on its entanglement dynamics.
More explicitely, the optimal control Hamiltonian that maximizes $\ddot \tau$ will steer the system towards states for which interactions create entanglement most rapidly, and for which entanglement is most robust against decoherence.
 
We can now determine the optimal control Hamiltonian in a straightforward fashion: since $\ddot \tau$ depends on $H_{\text{c}}$ linearly, this dependence can be phrased in terms of a scalar product
\be
\ddot\tau=\sum_i\frac{\partial\ddot\tau}{\partial h_i}h_i+\ddot\tau_0= \vec X \cdot \vec h + \ddot \tau_0
\label{eq:ddtau}
\ee
between the vectors $\vec h$ and $\vec X$.
The former contains the tunable parameters of the control Hamiltonian which, in the present case of local control, reads
$H_c=\sum_{ij}h_{i+3j}\sigma_i^j$, whith $\sigma_i^j$ the $i$-th Pauli matrix acting on the $j$-th spin.
$\vec X$ contains the partial derivatives of $\ddot\tau$ with respect to the $h_i$, which can be expressed algebraically in terms of the current system state, of the system Hamiltonian, and of the Lindbladian ${\cal L}$.
The additive term $\ddot \tau_0$ represents the free entanglement dynamics, in absence of control.
Given a maximal control intensity, the scalar product in Eq.~\eqref{eq:ddtau} and therefore the curvature $\ddot \tau$ is maximized by choosing $\vec h$ parallel to $\vec X$.
Following this procedure, we find the control Hamiltonian that is optimal for the instantaneous state of a given system in a purely algebraic fashion,
and its continuous application results in the desired time-dependent control Hamiltonian that induces an optimal entanglement evolution.

\begin{figure}
\centering
\includegraphics[width=0.6\columnwidth]{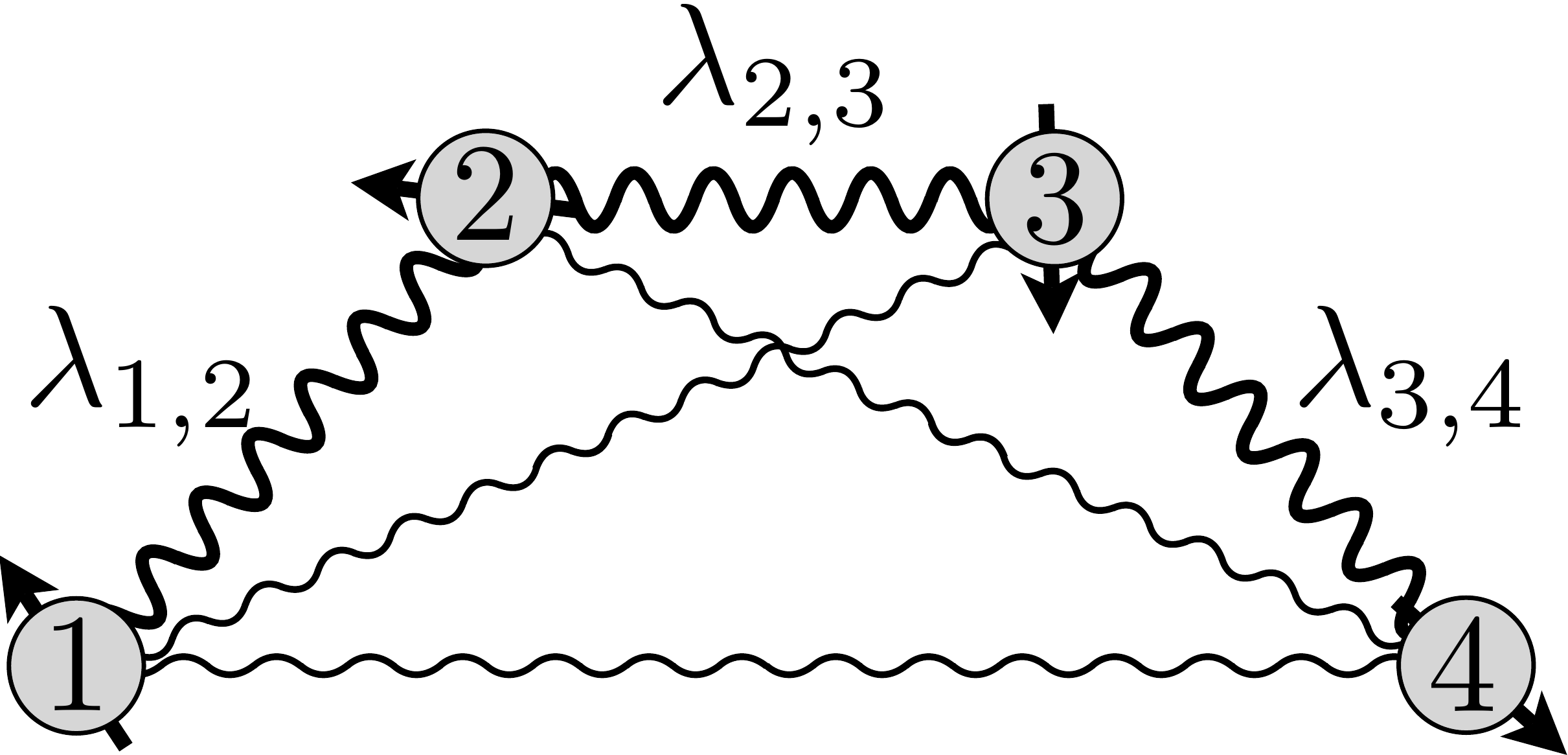} 
\caption{Four nitrogen vacancy (NV) centers in diamond, coupled by dipole-dipole interactions of different strengths. Numerical results obtained in this paper correspond to: $\lambda_{1,2} = 9.8$~MHz, $\lambda_{3,4} = 2.7$~MHz, and $\lambda_{2,3} = 1.3$~MHz, the largest three coupling constants, represented by thick lines.
Smaller coupling constants are represented by thin lines: $\lambda_{1,3} = 0.1$~MHz, $\lambda_{1,4} = 0.3$~MHz, and $\lambda_{2,4} = 0.5$~MHz.}
\label{fig:qubits}
\end{figure}

To be specific, we will now demonstrate the efficiency of our approach in a system of four nitrogen vacancy (NV) centers in diamond \cite{NVChildressLukin}, as sketched in FIG.~\ref{fig:qubits}.
The ground state electronic spin triplet of an NV center is a candidate for a quantum bit, since it can be manipulated via laser pulses \cite{NVmanip},
and read out non-destructively through fluorescence measurements \cite{NVreadout}. Furthermore, its coherence time, reaching up to a few micro seconds at room temperature, is only limited by weak coupling to the surrounding $^{13} C$ nuclear spin bath \cite{NVcohtime}.
All NV-centers in a diamond are coupled via the dipole-dipole interaction $H_{\text{sys}} = \sum_{i,j} \lambda_{i,j} \sigma_z^i \sigma_z^j$. The coupling constants $\lambda_{i,j}$ decay as $1/r^3$ with the distance $r$ between two centers. Since NV-centers have fixed positions, the $\lambda_{i,j}$ are not tunable, so that we have to resort to the aforementioned local control.

\begin{figure}
\centering
\includegraphics[width=\columnwidth]{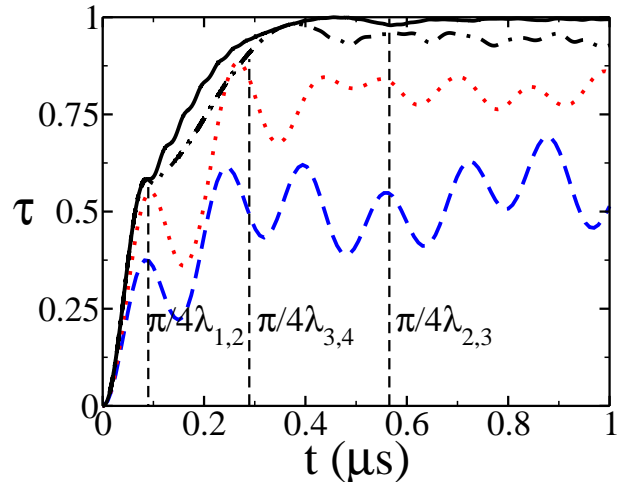} 
\caption{Time evolution of the lower bound of concurrence $\tau$ of a separable initial state of 4 spins coupled by dipole-dipole coupling as depicted in FIG.~\ref{fig:qubits}.
The free time evolution (dashed line), the time evolution under optimal local control with intensity $\lVert \vec h \rVert = 2.5$~MHz (dotted line), and $\lVert \vec h \rVert = 17$~MHz (solid line) are shown.
In addition, the dash-dotted line represents the time evolution of $\tau$ for completely unaddressable control pulses, with intensity $\lVert \vec h \rVert = 17$~MHz. Characteristic time-scales of $H_{\text{sys}}$ are indicated by dashed vertical lines.}
\label{fig:intensities}
\end{figure}

To start with, let us consider purely coherent dynamics.
The time evolution of $\tau$ is shown in FIG.~\ref{fig:intensities} for a separable initial state. If no control is applied (dashed line), entanglement oscillates around mediocre values on the characteristic time scales of $H_{\text{sys}}$ (indicated by dashed vertical lines), due to dipole-dipole interactions, and maximal entanglement is never reached.
In contrast, after switching on a weak local control, with maximal intensity $\lVert \vec h \rVert = 0.26 \, \lambda_{1,2}$, where $\lambda_{1,2}$ is the largest coupling constant, the system experiences a considerable increase of its average entanglement (dotted line). Yet, oscillations still prevail, since the dynamics induced by the control Hamiltonian is not fast enough to react to the natural system dynamics.
Nevertheless, by slightly increasing the control Hamiltonian's intensity, we can stabilize the amount of entanglement at ever higher average values, and an intensity of $1.7\, \lambda_{1,2}$ results in the evolution towards maximal entanglement (solid line). Oscillations of $\tau$ are flattened out, and once maximal entanglement is reached it is essentially maintained. Further augmenting $\lVert \vec h \rVert$ yields no substantial acceleration of entanglement creation, and saturation sets in. This is due to the fact that local control on its own {\em cannot} create any entanglement, and the only mechanism that does so is the dipole-dipole interaction which determines the minimal timescale $t_\text{E}$ on which maximal entanglement can be established. In general, at least $n-1$ pairwise couplings are needed to create genuine $n$-partite entanglement, and the smallest of these coupling constants determines the timescale of $t_\text{E}$. In the present case of four spins we have $t_\text{E} = \pi/4 \lambda_{2,3}$, where $\lambda_{2,3}$ is the third largest coupling constant
(FIG.~\ref{fig:qubits}).

Let us now explicitely analyze the dynamics induced by dipole-dipole interactions and optimal control according to Eq.~\eqref{eq:ddtau}. Since many different states exhibit the same amount of entanglement, the entanglement dynamics depicted in Fig.~\ref{fig:intensities} itself does not allow to infer the system's time evolution unambiguously.
Nevertheless, from the observation of the dynamics of many different separable initial states we can deduce that, first, spins $1$ and $2$ are entangled by their mutual interaction during an interval of $t = \pi/ 4\lambda_{1,2}$,
followed by the formation of a second entangled pair of spins $3$ and $4$, after $t = \pi / 4\lambda_{3,4}$.
Genuine four-particle entanglement and, hence, maximum entanglement is reached once the spins $2$ and $3$---from either one of the two already entangled pairs---have interacted for a time $t = \pi/4\lambda_{2,3}$.
More explicitly, this stepwise involvement of an increasing number of spins in the many-body entangled state reads
\begin{multline*}
	(\ket 0 + \ket 1)^{\otimes 4} 	\stackrel{\lambda_{1,2}}{\longrightarrow} (\ket {00} + \ket {11}) \otimes (\ket 0 + \ket 1)^{\otimes 2} \stackrel{\lambda_{3,4}}{\longrightarrow} \\
	(\ket {00} + \ket {11})^{\otimes 2} \stackrel{\lambda_{2,3}}{\longrightarrow} \ket{0000} + \ket{1111} + i(\ket{1100} + \ket{0011}),
\end{multline*}
where each arrow represents an interaction event, and all states are given up to local unitary transformations.

Compared to current NMR-techniques, the proposed control-scheme is much faster in creating entanglement. In~\cite{capluk}, an effective Hamiltonian $H_\text{eff}$ with uniform coupling between $n$ spins is implemented in the $L = n/2$ collective spin subspace, but in order to avoid leakage to other subspaces, the magnitude of the effective Hamiltonian has to be negligible with respect to the energy separation from other collective spin states. Therefore, the control Hamiltonian has to be chosen such that the effective uniform coupling is negligible with respect to the intrinsic dipole-dipole interaction---\ie\ $|H_\text{eff}|/|H_\text{sys}| = \epsilon \ll 1$. In the present case of 4 spins, $H_\text{eff}$ induces an evolution into a GHZ-state in $t_\text{GHZ}  = 32 \pi /  \epsilon \lambda$.
In our case, the creation of genuine multipartite entanglement is realized $128/\epsilon$ times faster, since there is no such limitation on $H_c$ and the full entangling capabilities of $H_{\text{sys}}$ are exploited.
Even the numerically powerful GRAPE-algorithm provides, at best, only very modest speed-up with respect to the entanglement creation rates achieved by our present control strategy:
an improvement by a factor of 0.9 is achieved for the creation of the symmetrically connected 4-qubit graph state,
while GRAPE yields no speed-up for the production of the 4-qubit 2D cluster state~\cite{fisher:042304}.

The above proves the effectiveness of our control strategy in the case of coherent dynamics. To test its reliability under more unfavourable circumstances, we want to impose additional restrictions.

First, it may be experimentally challenging to address all the spins individually, and, instead, a single driving field acts on all spins simultaneously.
In this case of {\em unaddressable} control, there are only the three components of the control Hamiltonian $H_c=\sum_{ij}h_i \, \sigma_i^{j}$ that can be adjusted.
Nevertheless, control with maximal intensity $\lVert \vec h \rVert = 1.7 \, \lambda_{1,2}$ (same as in the addressable case) results in the evolution into a highly entangled state as depicted in FIG.~\ref{fig:intensities} by a dash-dotted line. Even though the performance of these global control pulses is not as good as the performance of perfectly addressable pulses, it can be improved with higher maximum intensities, and an intensity of about $\lVert \vec h \rVert = 3 \, \lambda_{1,2}$ results in maximum values of $\tau$ (not shown here) that are comparable to those obtained when perfect addressability is given.

\begin{figure}
\centering
\includegraphics[width=\columnwidth]{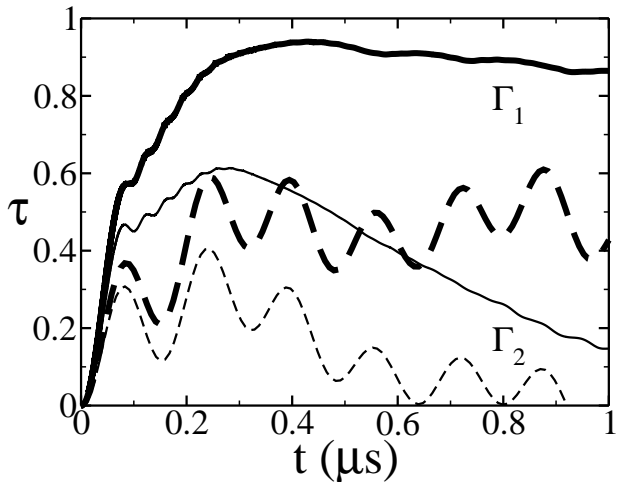} 
\caption{Time evolution of the lower bound of concurrence $\tau$ of the system depicted in FIG.~\ref{fig:qubits}, for a separable initial state under the influence of strong dephasing, at a rate $\Gamma_1=0.02 \, \mu s^{-1}$.
The uncontrolled dynamics is depicted by dashed lines, solid lines are obtained by optimal control with intensity $\lVert \vec h \rVert = 17$~MHz.}
\label{fig:diss}
\end{figure}

Second, unavoidable environmental effects render it necessary to consider optimal control of an open system which is subject to decoherence.
Here we focus on the paradigmatic model of dephasing---\ie\ $\mu_i = \sigma_z^i$ in Eq.~\eqref{eq:drho}.
The time evolution of $\tau$ in a system subject to dephasing at rate $\Gamma = 0.02 \, \mu \text{s}^{-1}$ is plotted in FIG.~\ref{fig:diss}, where the free dynamics in the absence of external control (as in Fig.~\ref{fig:intensities}) is depicted by a dashed line, and controlled dynamics is depicted by a solid line. Even in presence of this very strong decoherence---attainable coherence times in highly purified diamond are two orders of magnitude longer than $\Gamma^{-1}$ \cite{NVcohtime}---our control scheme performs very well, and control pulses with an intensity of $\lVert \vec h \rVert = 1.7 \lambda_{1,2}$ drive the system to almost maximal entanglement of $\tau = 0.95$.
Despite the fact that dephasing takes its toll on entanglement as time goes by, local control stabilizes the amount of entanglement at an average value
which almost doubles that of the uncontrolled system.
This is a consequence of our control strategy to maximize $\ddot\tau$ in Eq.~\eqref{eq:drho} that steers the system in a state least sensitive to the Lindblad term in Eq.~\eqref{eq:drho},
{\it i.e.} most robust against environment coupling.

A further aspect of the performance of control sequences in general is their robustness with respect to experimental imperfections such as intensity or phase fluctuations of the laser fields that are used to implement the control Hamiltonian.
Whereas a final assessment of this feature for our present scheme is beyond the scope of the present article, the construction of our control Hamiltonian
according to Eq.~\eqref{eq:ddtau} is insensitive to such errors at first order.
Indeed, tests with, {\it e.g.}, 10\% white noise added on top of a control pulse suggest a reduction of entanglement by 2\% only, in agreement with this expectation.

In conclusion, the invariance properties of an entanglement measure employed as a target functional---{\it i.e.} its independence of a single constituent's coherent dynamics---prove to establish significant advantages over prior approaches that optimize the target state fidelity.
Furthermore, by its very construction, our optimal control of the interplay of the system and environment dynamics provides new insight into the static and dynamical properties of quantum correlations in large and noisy composite quantum systems.
Finally, while demonstrated here with the help of concurrence, the validity of our present approach is by no means limited to this entanglement measure:
very specific types of entangled states, as for example required for the execution of a certain quantum information tasks \cite{onewayqc}
can be selected by target functionals formulated in terms of suitably chosen entanglement measures.

\bibliography{biblio}

\end{document}